# Study of the SmBaCuO solid solutions decomposition and its possible role for changing critical current

N.I. Matskevich, A.I. Romanenko, L.V. Yakovkina *(Institute of Inorganic Chemistry, Siberian Branch of the Russian Academy of Science, Russia)*, Th. Wolf *(Institute of Solid State Physics, Germany)*, G. Krabbes *(Institute of Solid State and Material Science, Germany)*

**Abstract.** We studied thermochemical characteristics of the $Sm_{1+x}Ba_{2-x}Cu_3O_y$ single crystals by solution calorimetry. Dependences of formation enthalpies from samarium content were constructed. It was established that solid solutions on the bases of Sm123 could be decomposed both in inert and in oxygen atmosphere into different mixtures. We supposed that solid solutions decomposition could lead to increasing critical current density. We assumed from thermochemical data that $J_c$ could be greater for samples prepared in oxygen than for samples synthesized in inert atmosphere. We confirmed these assumptions by comparison of obtained thermochemical data with transport properties measured in literature. We also investigated temperature dependences of resistance in the temperature range of 300–550 K during slow heating. As it was shown there was anomaly of resistance near 500 K. The origin of this anomaly was discussed.

## Introduction

The $REBa_2Cu_3O_y$ superconductors (RE123) are considered as promising materials for cryogenic applications. RE123 (RE = Nd, Sm, Eu, Gd) fabricated in the reduced oxygen atmosphere show a pronounced enhancement of critical current densities ($J_c$) in a few Tesla fields. A lot of papers concerning the Nd(Sm)–Ba–Cu–O systems are devoted to increasing critical current density and understanding origins of $J_c$ change. The consideration of published papers showed that peak effect was still contradictory. In this connection the investigations of physico-chemistry of the Sm–Ba–Cu–O system are important.

Authors [1–5] reached the highest values of critical current density for the $Nd_{1+x}Ba_{2-x}Cu_3O_y$ single phases. They studied some reasons of $J_c$ origin. There were reported in papers [1–4] that defects introduced from oxygen reduction, substitution or doping resulted in a peak effect with a large $J_c$ at higher fields. The maximum current depends on the kind of introduced defects, on twin structure and on field and temperature.

This paper is devoted to the physico-chemical study of the Sm–Ba–Cu–O system to get thermochemical data for the $Sm_{1+x}Ba_{2-x}Cu_3O_y$ phases. Then these data will compare with transport characteristics to understand some reasons of changing superconducting properties.

**Experimental**

We investigated two monocrystals. Stoichiometric Sm123 ($SmBa_2Cu_3O_{6.95}$) crystals were grown from CuO–BaO flux in an atmosphere of 300-mbar air in a $ZrO_2/Y$ crucible. They have been oxidized between 390–290° C during 300 h in 1 bar $O_2$. Their $T_c$ value is 94.5 K. $Sm_{1.11}Ba_{1.89}Cu_3O_{6.95}$ crystals have been grown in 1 bar air in $ZrO_2/Y$ crucible. The crystals have been oxidized in 1 bar $O_2$ between 600–400° C for about 650 h and show $T_c$ value of 77 K. All crystals were characterized by X–ray power diffraction and chemical analysis [6]. The content of copper was determined by atomic absorption spectrometry in the flame of air-acetylene, whereas the content of barium was determined by photometric method in the flame of nitrogen dioxiode-acetylene. The R.S.D. was 0.006–0.007. The content of samarium was determined by the spectrophotometry. The R.S.D. was 0.004–0.005. The stoichiometric coefficient of oxygen was determined by iodometric titration with an accuracy better than ±0.03. According to the results of the analyses the involved compounds were found to be single phases with an accuracy of about 1%.

Solution calorimetry was applied to measure thermochemical data. The experiments were performed in an automatic dissolution calorimeter with an isothermal shield. The calorimeter design and the experimental procedure are described in papers [7–8].

The 2$N$ HCl solution was used as a solvent. The experiments were performed at 323 K. Calorimetric cycles were designed in such a way that it was possible to determine the formation enthalpies of the Sm123 solid solutions from samarium oxide, copper oxide, and barium carbonate and then to calculate the formation enthalpies of $Sm_{1+x}Ba_{2-x}Cu_3O_y$ from oxides. The dissolution enthalpies were used to calculate the enthalpy of the following reaction ($\Delta_{car}H^0$):

$$0.5(1+x)\, Sm_2O_3 + (2-x)BaCO_3 + 3\, CuO + a\, O_2 = Sm_{1+x}Ba_{2-x}Cu_3O_y + (2-x)\, CO_2 \qquad (1)$$

On the basis of these data the formation enthalpy from binary oxides ($\Delta_{ox}H^0$) was calculated. The enthalpy of reaction $BaO + CO_2 = BaCO_3$ (– 272.4 kJ/mol [9]) was used for the calculation.

**Results and discussion**

We obtained the following data for the $Sm_{1+x}Ba_{2-x}Cu_3O_y$ single crystals: $\Delta_{ox}H^0$($SmBa_2Cu_3O_{6.95}$, s, 323.15 K) = –218.9±4.9 kJ/mol; $\Delta_{ox}H^0$($Sm_{1.11}Ba_{1.89}Cu_3O_{6.95}$, s, 323.15 K) = –160.0±5.3 kJ/mol.

In papers [10,11] we measured the formation enthalpies for barium cuprate and the ceramic Sm123ss (x = 0.4–0.8): $\Delta_{ox}H^0$($BaCuO_2$, s, 323.15 K) = –71.4±1.9 kJ/mol; $\Delta_{ox}H^0$($Sm_{1.4}Ba_{1.6}Cu_3O_{6.96}$, s, 323.15 K) = –153.3±5.9 kJ/mol; $\Delta_{ox}H^0$($Sm_{1.6}Ba_{1.4}Cu_3O_{7.16}$, s, 323.15 K) = –142.8±5.8 kJ/mol; $\Delta_{ox}H^0$($Sm_{1.8}Ba_{1.2}Cu_3O_{7.18}$, s, 323.15 K) = –118.6±5.1 kJ/mol.

Data obtained in this paper together with earlier measured enthalpies of ceramic $Sm_{1+x}Ba_{2-x}Cu_3O_y$ and barium cuprate enthalpy allowed one to construct dependences of $\Delta_{ox}H^0$ from x. Oxidation enthalpy equal to –96 kJ/mol O [12] was used to calculate formation enthalpies of solid solutions with y = 6 and y = 7. Graphs of obtained dependences are presented in Fig. 1.

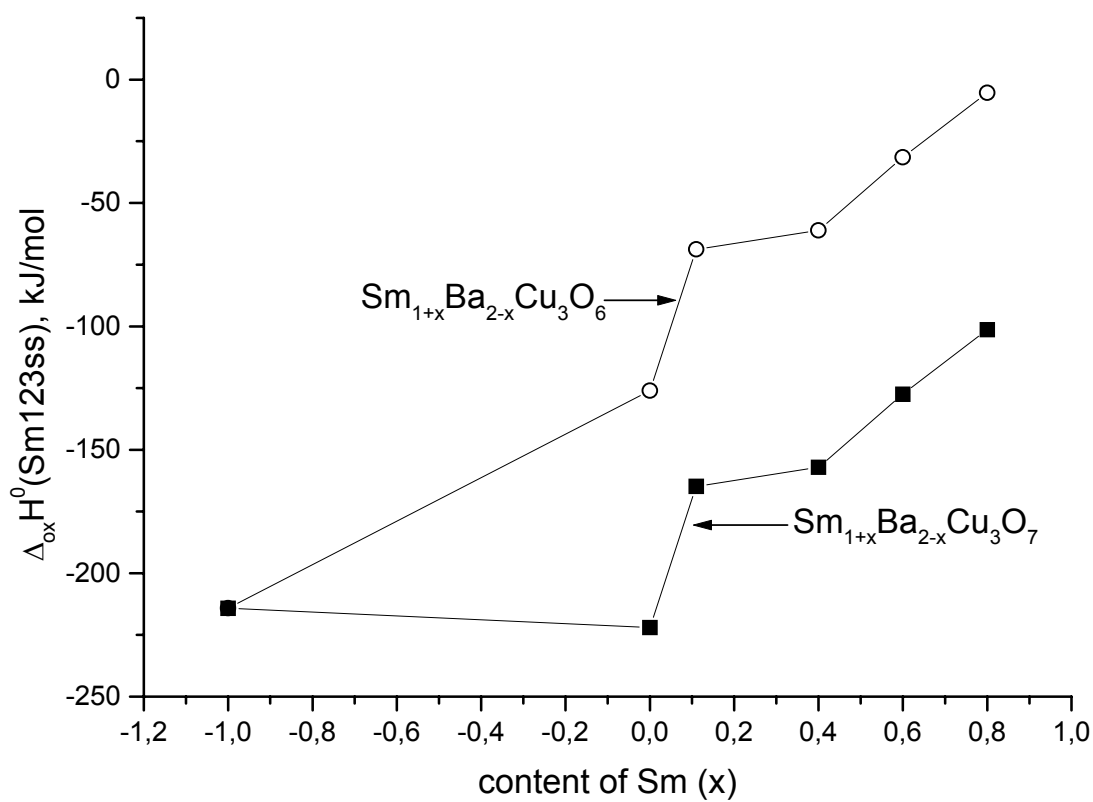

Fig. 1. Dependences of formation enthalpies of Sm123ss from Sm content

Consideration of obtained dependences allowed one to conclude that $Sm_{1+x}Ba_{2-x}Cu_3O_y$ could decompose into solid solutions with a high x+d and a low x–d and barium cuprate both in inert atmosphere and in oxygen atmosphere. I.e. the following reactions can proceed:

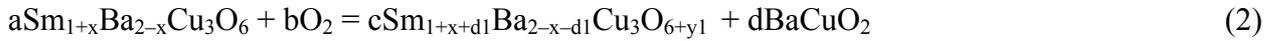
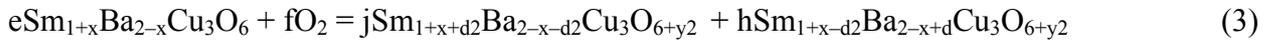

$aSm_{1+x}Ba_{2-x}Cu_3O_6 + bO_2 = cSm_{1+x+d1}Ba_{2-x-d1}Cu_3O_{6+y1} + dBaCuO_2$     (2)

$eSm_{1+x}Ba_{2-x}Cu_3O_6 + fO_2 = jSm_{1+x+d2}Ba_{2-x-d2}Cu_3O_{6+y2} + hSm_{1+x-d2}Ba_{2-x+d}Cu_3O_{6+y2}$     (3)

Then we will attempt to compare obtained thermochemical data with results on critical current density. Here it is necessary to say the following. Each type of pinning center, for example, RE211 inclusions, point defects, defect clusters, etc., has its own signature in the $J_c$ versus H dependence. It was shown that by processing RE123 in a low-pressure oxygen atmosphere, melt-textured samples with high critical currents in a pronounced peak could be obtained. This peak effect was attributed to the substitution of RE ions into Ba site, which could take place at decomposition of solid solutions and should act as a pinning center. As it was established, there were additional ways for decomposition in oxygen in comparison with inert atmosphere. For this reason it is possible to assume that critical current density of samples annealed in oxygen atmosphere will be greater than of samples prepared in inert gas. We will try to confirm this conclusion based on data of paper [13].

In paper [13] the influence of several conditions for post-annealing on superconducting properties has been investigated in the SmBaCuO/Ag composites. It has been shown that samples of SmBaCuO/Ag with high superconducting characteristics can be prepared in air and their properties can be improved by a post-annealing under low oxygen partial pressure. Samples were post-annealed at 925° C in 0.1% $O_2$. The significant improvement in the critical current densities and irreversibility fields is observed. Critical current density was increased from $1.5 \cdot 10^4$ A/(sm$^2$) up to $2.8 \cdot 10^4$ A/(sm$^2$) after post-annealing in 0.1% $O_2$. So, obtained thermochemical data are confirmed by results on measurements of HTSC transport characteristics. Confirmation of decomposition is given in paper [14] for the Nd–Ba–Cu–O system. Using Raman spectroscopy it was shown that barium cuprate is formed at high temperature. Furthermore decomposition of solid solutions existence was confirmed by electron microscopy [15] also for NdBaCuO. It is possible to believe that a similar confirmation of decomposition is true for $Sm_{1+x}Ba_{2-x}Cu_3O_y$ as well.

For understanding the change of superconducting characteristics, it is also useful to study anomalies of physico-chemical properties at high temperature. We investigated the temperature dependences of resistance in the temperature range of 300–550 K during slow heating (≤ 1K/h). Procedure and technique of experiments were described in papers [16–18]. We showed that there was anomaly of resistance near 500 K (see, fig. 2). It is necessary to note that linear temperature dependence of electroresistance is usually observed $\rho(T) \propto T$. The SmBaCu$_3$O$_y$ polycrystal was also studied by means of differential scanning calorimetry and differential thermal analysis. It was found that at about 470 K the polycrystal sample shows an anomaly without a weight loss. This anomaly was found for the Sm123 sample for the first time.

The analogous anomaly was found for the Y123 phase by differential scanning calorimetry [19] and drop calorimetry [20]. There are different opinions about origin of this anomaly. These opinions are presented in papers [20]. But it is undoubtedly necessary to take into account this anomaly at synthesis of HTSC.

As consideration of literature showed, HTSC samples were always annealed at 300–600 K. Taking into account information about the anomaly existence it is possible to recommend two different temperature for annealing samples: 1) annealing samples at temperature higher than 500 K; 2) annealing samples at temperature lower than 500 K. May be it will give the possibility to get HTSC cuprates with new properties.

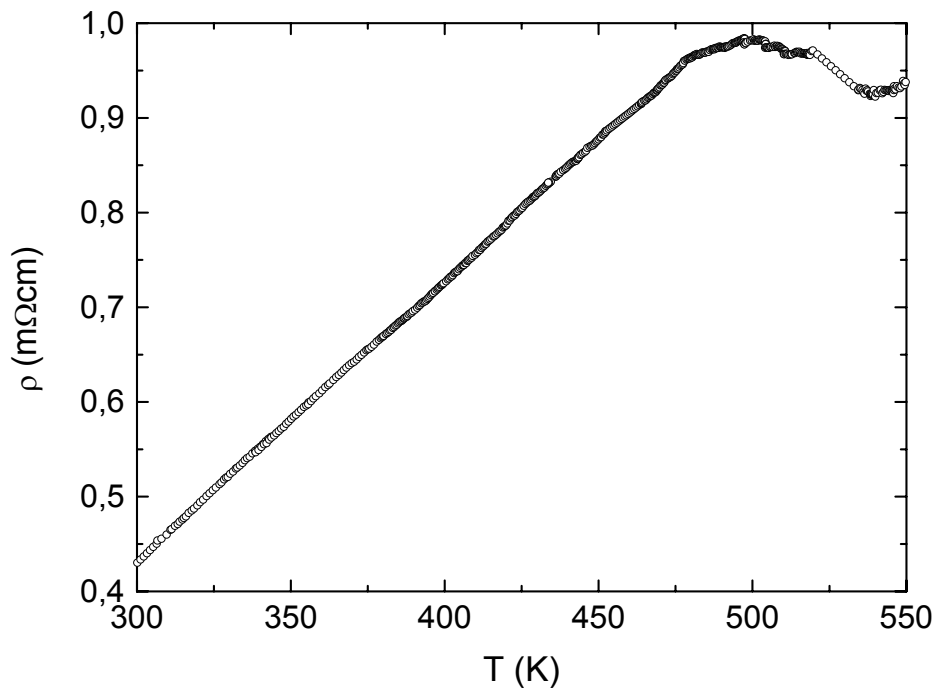

Fig. 2. Temperature dependence of specific electroresistence (resistivity) $\rho(T)$ of single crystal SmBa$_2$Cu$_3$O$_y$ at slow heating ($\leq$ 1K/h)

**Conclusion**

1. Thermochemical characteristics of the $Sm_{1+x}Ba_{2-x}Cu_3O_y$ single crystals were studied. Dependences of formation enthalpies from samarium content were constructed.

2. It was established that solid solutions on the bases of Sm123 could be decomposed both in inert and in oxygen atmosphere into different mixtures.

3. Obtained thermochemical characteristics were compared with transport properties. It was supposed that decompositions of solid solutions could lead to increasing critical current density. It was possible to assume from thermochemical data that $J_c$ could be greater for samples prepared in oxygen than for samples synthesized in inert atmosphere. This assuming was confirmed by transport characteristics measurement.

4. Temperature dependences of resistance in the temperature range of 300–550 K during slow heating were investigated. As it was shown, there was anomaly of resistance near 500 K.

**Acknowledgements**

This work is supported by Russian Foundation for Basic Research (Project No 02–03–32514) and contract of Karsruhe Investigated Center (Grant N 0001/00972158/5140).